\documentclass[12pt]{iopart}

\usepackage[separate-uncertainty=true, multi-part-units=single]{siunitx} 
\usepackage{braket}		
\usepackage{graphicx}
\usepackage{tabularx}				
\usepackage{xcolor,colortbl}       

\begin{document}

\title[Sagnac-type entangled photon source using only conventional polarization optics]{Sagnac-type entangled photon source using only conventional polarization optics}

\author{Youn Seok Lee$^{1}$, Mengyu Xie$^{1,2}$, Ramy Tannous$^{1}$ and Thomas Jennewein$^{1}$}

\address{$^1$Institute for Quantum Computing and Department of Physics and Astronomy, University of Waterloo, Waterloo, Ontario N2L 3G1, Canada}
\address{$^2$State Key Laboratory of Low-Dimensional Quantum Physics and Department of Physics, Tsinghua University, Bejing 100084, China}
\ead{ys25lee@uwaterloo.ca}
\vspace{10pt}
\begin{indented}
\item[]June 2020
\end{indented}

\begin{abstract}
We designed and implemented a novel combination of a Sagnac-interferometer with a Mach-Zehnder interferometer for a source of polarization-entangled photons. The new versatile configuration does not require multi-wavelength polarization optics, yet it performs with a good polarization quality and phase-stability over a wide wavelength range. We demonstrate the interferometer using only standard commercial optics to experimentally realize the pulsed generation of polarization-entangled photon-pairs at wavelengths of \SI{764}{\nm} and \SI{1221}{\nm} via type-I spontaneous four-wave mixing in a polarization-maintaining fiber. Polarization entanglement was verified by a polarization-correlation measurement with a visibility of \SI{95.5}{\percent} from raw coincidence counts and the violation of the Clauser-Horne-Shimony-Holt (CHSH) inequality with $S=2.70\pm0.04$. The long-term phase-stability was characterized by an Allan deviation of \SI{8}{\degree} over an integration time of about 1 hour with no active phase-stabilization.
\end{abstract}

%
\vspace{2pc}

\section{Introduction}\label{sec:intro}

Polarization-entangled photons have been a crucial element in the quantum optics community for fundamental tests of quantum mechanics, quantum communication, as well as quantum metrology~\cite{QuantumEntanglement}. In particular, with the recent developments in satellite-based quantum communication~\cite{Satellite_PKnight, Satellite_AZeilinger, Satellite_ALing}, entangled photon-pairs at widely separated wavelengths are desirable for the direct entanglement between ground-based fiber optical networks and space. The photon transmission for free-space quantum channels is optimal at near-infrared (\SIrange{780}{810}{\nano\meter})~\cite{Compre_design_JP} wavelengths, whereas telecom wavelengths (\SIrange{1310}{1550}{\nano\meter}) are optimal for optical fiber channels. These non-degenerate polarization-entangled photon sources have been implemented with various nonlinear materials and schemes, such as a dispersion-shifted fiber in Michelson-interferometer~\cite{FiberMichelson_PremKumar}, periodically poled potassium titanyl phosphate in a so-called sandwich configuration \cite{Sandwitch_AKarlsson, Sandwitch_AZeilinger} and in a Sagnac-interferometer~\cite{SagnacPeriscope_AZeilinger}, and periodically poled lithium niobate in a Sagnac-interferometer~\cite{UnfoldedSagnac_AKarlsson}.

Sagnac-type entangled photon sources (Sagnac-EPS) have been one of the most popular methods due to their intrinsic phase-stability and versatility~\cite{Sagnac_Kim} . A variety of nonlinear optical media including nonlinear optical fibers, photonic crystal fibers, and bulk- and waveguide-type periodically poled nonlinear crystals have been employed~\cite{DispersionShiftedFiberSagnac, CommercialFiberSagnac, MicrostructureFiberSagnac, PhotonicCrystalFiberSagnac, PPLN-WG_Sagnac, PPLN-ridgeWG_Sagnac, PPLN-WG_Sagnac_fullfiber, PPKTP-WG_Sagnac}. However, a typical Sagnac-EPS requires polarization optics working in at least two, sometimes three, very distinct wavelengths, i.e., pump and photon-pairs. The customization of polarization optics for highly non-degenerate photon-pairs is technically demanding and costly, and its performance typically limits the quality of the polarization-entanglement. Sauge \etal~\cite{UnfoldedSagnac_AKarlsson} demonstrated a non-degenerate entangled photon source by unfolding the Sagnac-loop into two different loops: one for pump (\SI{532}{\nano\meter}) and idler (\SI{1550}{\nano\meter}) photons and the other for signal (\SI{810}{\nano\meter}) photon. This unfolded Sagnac scheme, which is originally proposed by Fiorentino \etal~\cite{Constraint_released_Sagnac}, provides flexibility in the choice of the wavelength without the customization of polarization optics. Recently, the unfolded Sagnac-like entangled photon source was implemented for the randomness extraction from Bell violation~\cite{Kurtisefer_RandomnessExtraction}. However, the scheme loses the intrinsic phase-stability of an original Sagnac-EPS due to the separate optical paths for the signal and idler photons. On the other hand, Hentschel \etal~\cite{SagnacPeriscope_AZeilinger} demonstrated an original Sagnac-EPS by replacing a polarization beam splitter and a half-waveplate with a Glan-Thompson polarizer and a periscope, respectively. In their setup, a custom-designed Glan-Thompson polarizer is required to avoid angular dispersions and the optical alignment is relatively tedious. 

Mach-Zehnder-type entangled photon sources (MZ-EPS) with two beam displacers have recently become popular~\cite{BD_Beausoleil, BD_THumble, LoopholeFreeTest, BD_RolfHorn, SingaporeSource_ALing}. The main advantages of this design are the simple alignment procedure and the compactness, while the monolithic configuration exhibits a high phase-stability. In addition, the MZ-EPS takes full advantage of the high polarization extinction ratio of beam displacers over a wide range of wavelengths. However, since the photon-pair production processes occur in two different optical paths, one may have to prepare two identical nonlinear materials or miniaturize the waveplates used. More importantly, the MZ-EPS cannot be easily adapted for fiber- or waveguide-based nonlinear media due to the requirement of their guided modes to be matched.

In this report, we develop a new interferometer for polarization-entangled photon sources which is suitable for various optical nonlinear materials over a wide wavelength range without the need for customized polarization optics. The interferometer is configured such that a Sagnac-loop is placed inside a Mach-Zehnder interferometer that is formed by two beam displacers. The polarization states of the pump and photon-pairs are split and recombined by two beam displacers as in MZ-EPS, and at the same time, two pair-production processes are kept in a common optical path as in Sagnac-EPS. Thus, the designed entangled photon source, which we call a beam displacement Sagnac-type entangled photon sources (BD-Sagnac-EPS), takes advantage of both an original Sagnac-EPS and MZ-EPS. We demonstrate the designed interferometer with the pulsed generation of polarization-entangled photon-pairs at the wavelengths of \SI{764}{\nm} and \SI{1221}{\nm} via spontaneous four-wave mixing (SFWM) in a commercial-grade polarization-maintaining fiber (PMF) by using only standard commercial optical elements. 

\begin{figure}[t]
\centering
\includegraphics[width=10cm]{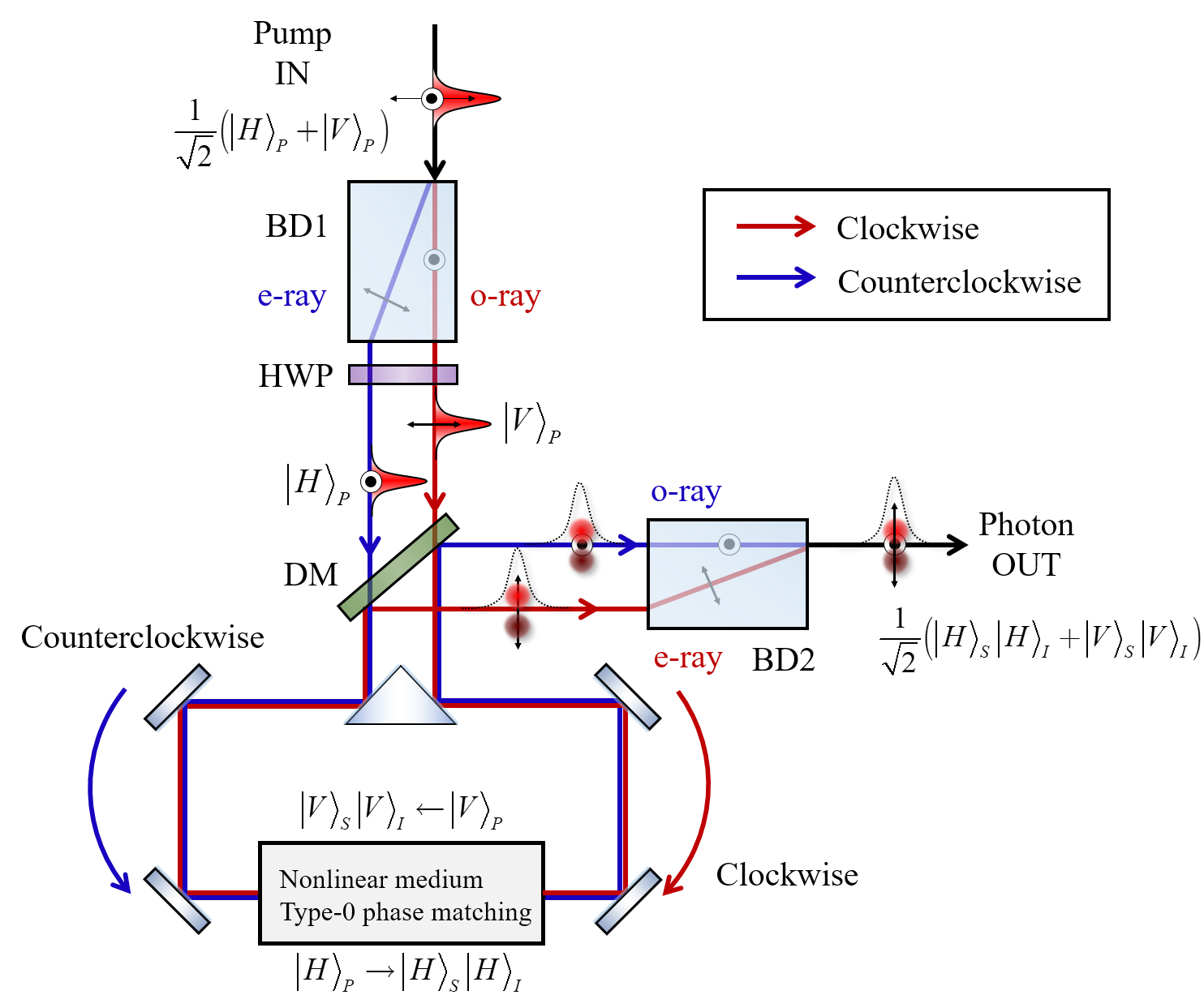}
\caption{A schematic drawing of the conceptual design for BD-Sagnac-EPS; BD, beam displacers; HWP, half-waveplate; DM, dichroic mirror; o-ray, ordinary ray; e-ray, extraordinary ray. Here, we consider the type-0 phase-matching condition as an example of the pair generation process. The implementations for type-1 and type-2 phase-matching conditions are discussed in the main text.}
\label{fig:concept_design}
\end{figure}

\section{Conceptual design}\label{sec:concept_design}
The conceptual design of the BD-Sagnac-EPS is depicted in figure~\ref{fig:concept_design}. Two beam displacers (BD1 and BD2) form a Mach-Zehnder interferometer; one (BD1) is for the input port of pump light and the other (BD2) is for the output port of the generated photon-pairs. The diagonally polarized pump light enters the interferometer, and the BD1 converts the two orthogonal polarizations into two parallel optical paths, $|H\rangle_{P}$ and $|V\rangle_{P}$. By connecting one path to the other, the Sagnac-loop is formed inside the Mach-Zehnder interferometer (blue path: clockwise and red path: counterclockwise). The optical nonlinear medium is placed at the center of the Sagnac-loop where the correlated photon-pairs (namely, signal and idler) are generated via a spontaneous parametric amplification process, e.g., a spontaneous parametric down-conversion and four-wave mixing. As an example, if we take type-0 phase-matching conditions as the pair production process we obtain $|V\rangle_{P} \rightarrow |V\rangle_{S}|V\rangle_{I}$ (clockwise) and  $|H\rangle_{P} \rightarrow |H\rangle_{S}|H\rangle_{I}$ (counterclockwise). Here, the subscripts $P$, $S$, and $I$ stand for the pump, signal, and idler photons, respectively. As the wavelengths of the generated photons are different from that of the pump, the two orthogonally polarized photon-pairs are spectrally filtered out using a dichroic mirror which in our case was implemented with a standard bandpass filter, and exit the Sagnac-loop. \footnote{This could also be realized by a standard notch filter centered at the pump's wavelength. In this case, the incident pump light is reflected and the generated photon-pairs are transmitted. Note that a precise spectral filter or dichroic mirror is not required because the pump wavelength is usually far apart from the photon-pair's wavelengths in parametric down-conversion and four-wave mixing process.} Then, the BD2 recombines the two photon-pairs into a single spatial mode. The superposition of two pair-production processes yields the polarization-entangled state, $\ket{\Phi^+}=\left(\ket{V}_{S}\ket{V}_{I} + \ket{H}_{S}\ket{H}_{I}\right)/\sqrt{2}$.

Beam displacers are made of uniaxial birefringent materials where refractive indices between ordinary and extraordinary rays are different. The refractive index discrepancy for a few centimeter beam displacer yields an optical path length difference that is large enough to timely separate the photons' wavepackets of two polarization modes, i.e., a temporal walk-off. To make the optical path lengths balanced within the interferometer, the temporal walk-off caused by the BD1 must be compensated at the BD2. In this example of the type-0 phase-matching condition, a half-waveplate is introduced after the first beam displacer in order to flip the pump polarization such that ordinary/extraordinary ray at the first beam displacer experiences the extraordinary/ordinary path at the second beam displacer. Note that for the type-1 phase-matching condition, the interferometer is balanced without the half-waveplate in figure~\ref{fig:concept_design}.

\begin{figure}[t]
			\includegraphics[width=0.75\columnwidth]{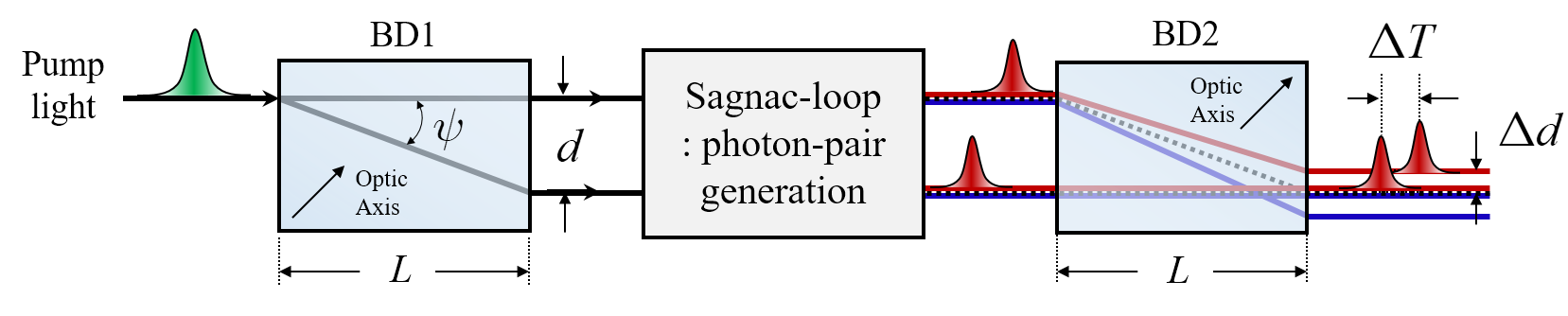}
			\centering
			\caption{Spatial and temporal walk-off due to the chromatic dispersion of beam displacers (BD1 and BD2). The photons generated from an optical nonlinear medium in the Sagnac-loop have different wavelengths from the incident pump light, and therefore they pass through the BD2 with different refractive angles and group velocities.}
			\label{fig:walk-off_EPS}
\end{figure}

However, for non-degenerate entangled photon sources, the three photons, i.e., pump, signal, and idler, for each polarization mode propagate through the beam displacer at different speeds and refraction angles due to its chromatic dispersion. This causes additional spatial and temporal walk-offs that must be compensated before or after the interferometer appropriately. As illustrated in figure~\ref{fig:walk-off_EPS}, the spatial walk-off decreases the fiber-coupling efficiency of the entangled photon-pairs and the temporal walk-off degrades the entanglement quality due to the imperfect overlap between the photon wavepackets in the ordinary and extraordinary paths~\cite{LoudonBook}. Furthermore, the phase-shifts caused by the temperature variation are different for the three photons due to the chromatic dispersion, which may impact the stability of the interferometer significantly. In the following, we estimate the walk-offs and the temperature-dependent phase-shift of the entangled state, for different uniaxial birefringent crystals.

First, we estimate the spatial walk-off $\Delta d_{S(I)}$ by calculating the refraction angles $\psi$ between ordinary and extraordinary rays in the beam displacers~\cite{BDformula}, as expressed by
\begin{equation}
\tan\left(\psi\right) = \frac{\left(n_e^{2}-n_o^{2}\right)\cos(\theta)\sin(\theta)}{n_e^{2}\cos^{2}(\theta)+n_o^{2}\sin^{2}(\theta)},
\label{Beamdisplacer}
\end{equation}
where $\theta$ is the angle between the optical axis and the wavevector of the incident light. Here, $n_o$ and $n_e$ are denoted as the ordinary and extraordinary refractive indices, respectively. From the Sellmeier equation for different birefringent materials, the spatial walk-off is calculated for the optical axis aligned at the angle of \SI{45}{\degree} with respect to the wavevector of incident light: 
\begin{equation}
\Delta d_{S(I)} = L \tan\left(\psi_{P}\right) - L \tan\left(\psi_{S(I)}\right),
\end{equation}
where $L$ is the length of beam displacers and $\psi_{P(S,I)}$ is the refraction angle of the pump, signal, and idler photons, respectively.

Secondly, to estimate the temporal walk-off $\Delta T_{S(I)}$, we consider the arrival time of the photon-pairs at the end of the BD2 after traveling through the ordinary and extraordinary paths of the interferometer. These arrival times for signal and idler photons are referenced from when the pump pulse departed the front of the BD1 and can be expressed as
\begin{equation}
T_{\text{S(I)},o(e)} = \frac{l_{e(o)}(\lambda_{\text{P}})}{v_{g,e(o)}(\lambda_{\text{P}})} + \frac{l_{o(e)}(\lambda_{\text{S(I)}})}{v_{g,o(e)}(\lambda_{\text{S(I)}})},
\label{ArrivalTime}
\end{equation}
where $l_{e(o)} = L/\cos(\psi)$ is the physical optical path length for the extraordinary (ordinary) light of the beam displacer and $v_{g,e(o)}(\lambda_{P(S,I)}$) is the group velocity of the extraordinary (ordinary) light at the wavelength of the pump, signal, and idler photons, respectively. Note that the group velocity for the extraordinary ray was calculated from the effective refractive index $n_{\text{eff}}$, defined as
\begin{equation}\label{eq:eff_ne}
\frac{1}{n_{\text{eff}}^2}=\frac{\sin^{2}(\theta_{e})}{n_{e}^{2}} + \frac{\cos^{2}(\theta_{e})}{n_{o}^{2}},
\end{equation}
where $\theta_{e}=\psi+\theta$ is the angle between the optic axis and the wavevector of the extraordinary ray. Then, the temporal walk-off can be calculated by the arrival time difference between ordinary and extraordinary light,
\begin{equation}
\Delta T_{S(I)} = T_{\text{S(I)},o} - T_{\text{S(I)},e}.
\end{equation}

Finally, the temperature-dependent phase-shift is estimated for the entangled state which can be written as 
\begin{equation}
\ket{\Phi^+}=\frac{1}{\sqrt{2}}\left(\ket{V}_{S}\ket{V}_{I} + e^{i\phi(\lambda,T)}\ket{H}_{S}\ket{H}_{I}\right).
\end{equation}
Here, the $\phi(\lambda,T)$ is the relative phase between the clockwise and counterclockwise optical paths. As for the optical elements that lie inside the Sagnac-loop, the temperature-dependent phase-shift can be negligible due to the self-compensation effect~\cite{SagnacPeriscope_AZeilinger} and we ignore the phase-shift due to the waveplate after the first beam displacer because of its thickness. Then, the relative phase can be expressed as
\begin{eqnarray}
\phi(\lambda,T) &= 2\phi_{e}(\lambda_P) + \phi_{o}(\lambda_S) + \phi_{o}(\lambda_I) - \left(2\phi_{o}(\lambda_P) + \phi_{e}(\lambda_S) + \phi_{e}(\lambda_I)\right)\\
&\equiv 2\Delta\phi(\lambda_P) - \Delta\phi(\lambda_S) - \Delta\phi(\lambda_I).
\end{eqnarray}
Here, the factor of two for the pump's phase-shift comes from the fact that in the four-wave mixing process two pump photons are converted to the signal and idler photons. It is worth to note that the phase-shift difference between the ordinary and extraordinary rays $\Delta\phi(\lambda) = \phi_{e}(\lambda) - \phi_{o}(\lambda)$ for signal and idler photons are compensated by the pump-photon. This is the manifestation of the self-compensating effect of the Saganc-interferometer~\cite{SagnacPeriscope_AZeilinger}. In fact, our designed interferometer can be viewed as an unfolded Sagnac-interferometer: a polarized beam splitter is replaced by two beam displacers, and it maintains the inherent stability of the Sagnac-interferometer. The temperature-dependent phase-shift is characterized by the derivative of the phase-shift with respect to the temperature $\partial\phi/\partial T$. Each term can be expressed in terms of the thermo-optic coefficient and the thermal expansion coefficient as
\begin{equation}
\frac{\partial\phi}{\partial T} = \frac{2\pi}{\lambda}L(T)\bigg[\frac{\partial n(\lambda,T)}{\partial T}+n(\lambda,T)\alpha\bigg],
\end{equation}
where $\alpha$ is a thermal expansion coefficient. Again, the thermo-optic coefficient for the extraordinary ray is derived from the formula of the effective refractive index~\ref{eq:eff_ne}.
\begin{equation}
\frac{\partial n}{\partial T} = \frac{\frac{\partial n_{o}}{\partial T}n_{e}+n_{o}\frac{\partial n_{e}}{\partial T}}{\sqrt{n_{o}^2\sin^{2}(\theta_{e}) + n_{e}^2\cos^{2}(\theta_{e})}} + n_{o}n_{e}\frac{2n_{o}\frac{\partial n_{o}}{\partial T}\sin^{2}(\theta_{e}) + 2n_{e}\frac{\partial n_{e}}{\partial T}\cos^{2}(\theta_{e})}{(n_{o}^2\sin^{2}(\theta_{e}) + n_{e}^2\cos^{2}(\theta_{e}))^{3/2}}
\end{equation}
It is worth noting that one can design the beam displacers by cascading two or more complementary birefringent materials with the carefully chosen length ratio such that the first-order temperature-dependence of the phase-shift vanishes~\cite{book_PolarizationOptics}. However, in our analysis, we restrict our scope to the two beam displacers with the same length and birefringent material for the practical implementation of the BD-Sagnac-EPS. 

We demonstrate our design by generating photon-pairs at \SI{764}{\nano\meter} and \SI{1221}{\nano\meter} via SFWM driven by the pump light at \SI{940}{\nano\meter}. The spatial $\Delta T_{I}$, temporal walk-off $\Delta T_{S}$, and the temperature-dependent phase-shift $\partial\phi/\partial T$ are calculated for three common birefringent materials, i.e., calcite, $\alpha$-BBO, and YVO$_{4}$, for \SI{40}{\milli\meter}--long beam displacers at the wavelengths of the photon-pairs. The results are summarized in Table.~\ref{table:Walk-Off}. We found that the temperature-dependence of the phase-shift is less than $0.04\pi/$K for the three materials and the YVO$_{4}$ is the most promising candidate in terms of thermal stability. As the temporal walk-off is relatively easy to compensate by birefringent materials in comparison with the spatial walk-off, the $\alpha$-BBOs is also a good choice. However, in our demonstration, we used two calcite beam displacers due to the availability of these components at our facilities. 

\begin{table}[htbp]
\caption{\label{table:Walk-Off}The spatial, temporal walk-off, and temperature-dependent phase-shift caused by chromatic dispersions from different materials for two beam displacers in BD-Sagnac-EPS. The Sellmeier equations $n_{o(e)}$, thermo-optic coefficient $dn/dT$, and thermal expansion coefficient $\alpha$ for calcite and $\alpha$-BBO are obtained from~\cite{book_Ghosh}. Those parameters for YVO$_{4}$ were from~\cite{thermo_optic_YVO4}. Note that in~\cite{thermo_optic_YVO4} the thermo-optic and thermal expansion coefficients are available at the wavelength of \SI{0.9}{\micro\meter}, \SI{1.1}{\micro\meter}, and \SI{1.3}{\micro\meter}. We applied a linear interpolation and extrapolate to obtain the values at the wavelength of \SI{764}{\nano\meter}, \SI{940}{\nano\meter}, and \SI{1221}{\nano\meter}.}
\begin{indented}
\item[]\begin{tabular}{@{}lllll }
\br
Material & $\Delta T_{S}, \Delta T_{I}$ (\si{\pico\second}) &   	$\Delta d_{S}, \Delta d_{I}$ (\si{\milli\meter})	&   	$\partial\phi/\partial T$	(\si{\degree}/K)\\
\mr
 Calcite	& -0.20, 0.06			&	-0.07, 0.09	&	-7.00	\\
$\alpha$-BBO	  & -0.18, 0.15 	&	-0.01, -0.01	&	-0.97	\\
YVO$_{4}$  & 1.35, -0.93		&	0.07, -0.06	&	-0.86\\
\br
\end{tabular}
\end{indented}
\end{table}

After accounting for the spatial and temporal walk-off, as well as the temperature-dependent phase-shift, here we summarize the main advantages of the BD-Sagnac-EPS. First, the setup does not require any customized polarization optics because pump and photon-pairs do not share any common polarization optics. In fact, our demonstration consists of all commercial off-the-shelf optical elements. Protected-silver coated mirrors have a nominal reflectivity of greater than \SI{97}{\percent} for wavelengths between \SIrange{0.450}{7}{\micro\meter} while the two identical calcite beam displacers exhibit extinction ratios of 100,000:1 over wavelengths between \SIrange{0.200}{3.5}{\micro\meter}, which makes the setup suitable for highly non-degenerate polarization-entangled photon-pairs. Secondly, the interferometer is symmetric and the optical path lengths are balanced for two orthogonal polarization states by the Sagnac-loop, and therefore pulsed operation can be readily achieved. Third, the alignment procedure is easier than for typical Sagnac-type sources due to the fact that the optical path of the incident pump light and the emitted photon-pairs can be individually controlled, meaning that the optical paths for the pump and photon-pairs do not have to be on the same plane. Finally, our design is applicable for as broad range of optical nonlinear media as with typical Sagnac-EPS, including an optical nonlinear bulk crystal~\footnote{One may place two identical nonlinear bulk crystals that are cross-oriented to each other~\cite{two-crystal-Sagnac}.}, nonlinear optical fibers, and periodically poled nonlinear waveguides. Furthermore, our design is not limited to the type-0 phase-matching condition described above, as type-1 and type-2 phase-matching conditions can also be implemented with minor modifications. For example, the type-1 condition can be easily implemented by removing the half-waveplate. For the type-2 condition, one may add one or two more beam displacers to rearrange the polarization components at the output, as demonstrated in other works~\cite{LoopholeFreeTest, BD_RolfHorn}. 

\begin{figure}[t]
\centering
\includegraphics[width=10cm]{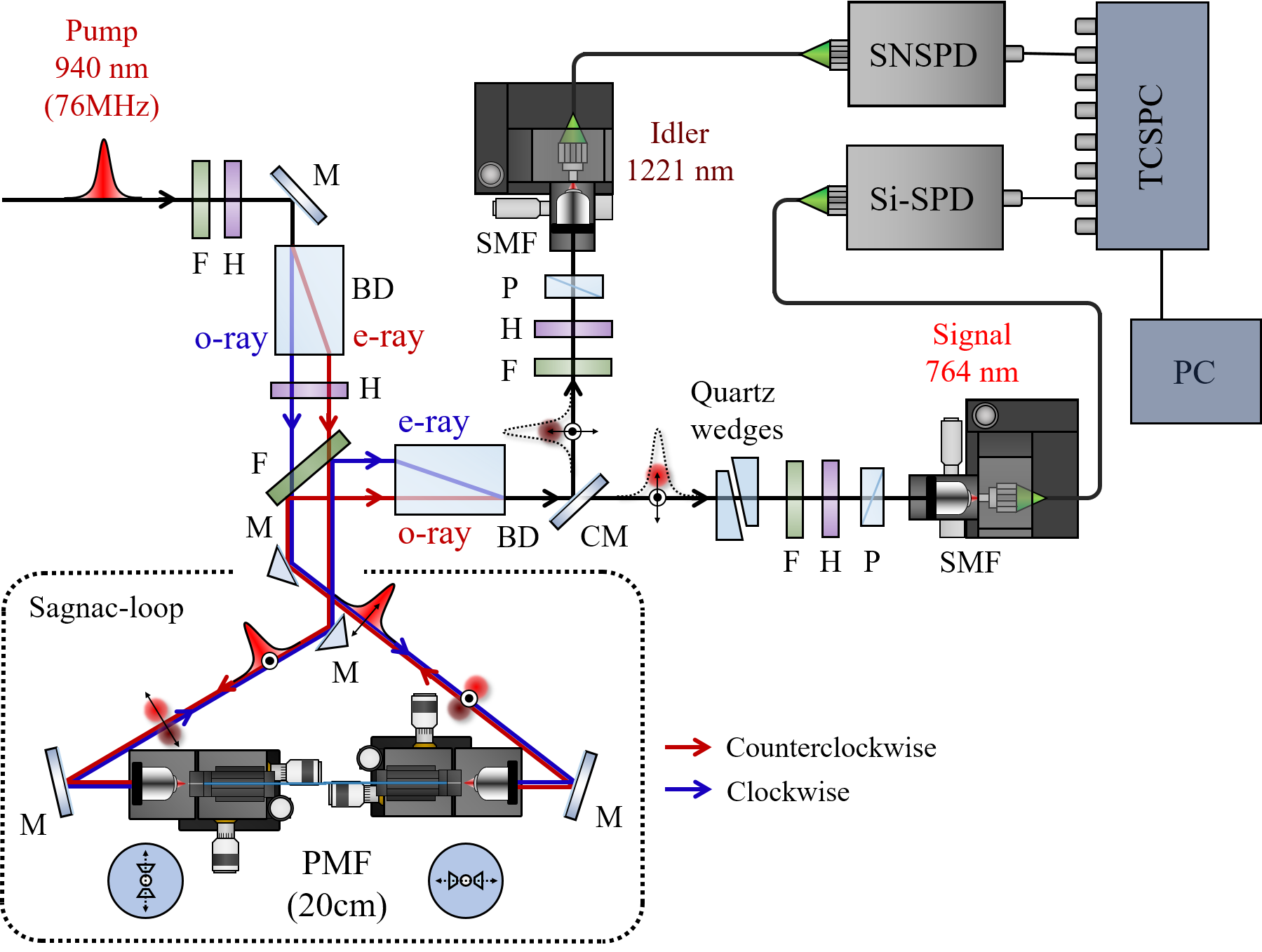}
\caption{A schematic drawing of the experimental setup for the demonstration of the new interferometer design. Correlated photon-pairs are generated from \SI{20}{\centi\meter}-long PMF that is pumped by mode-locked picosecond pulse laser (\SI{940}{\nano\meter}, \SI{76}{\mega\hertz}, \SI{3}{\pico\second}). The PMF is placed at the center of Sagnac-loop and twisted by \SI{90}{\degree}, such that the slow-axis is aligned to pump polarizations. The generated photon-pairs exit the Sagnac-loop upon the reflection by a spectral bandpass filter; BD, \SI{40}{\milli\meter}-long calcite beam displacer; F, bandpass filter; H, half-wave plate;  P, linear polarizer; M, protected-silver-coated mirror; DM, dichroic mirror;  SMF, single-mode fiber; SNSPD, superconducting nanowire single-photon detector; Si-SPD, Silicon-based single-photon detector; TCSPC, time-correlated single-photon counter.}
\label{fig:ExpSetup}
\end{figure}

\section{Experimental setup}

We experimentally implement our design and generate non-degenerate polarization-entangled photon-pairs via SFWM in a commercially available PMF (HB800G, Thorlabs). A schematic of the experimental setup is presented in figure~\ref{fig:ExpSetup}. Diagonally polarized pump light from a mode-locked laser, operating at a wavelength of \SI{940}{\nano\meter} with a pulse duration of \SI{3}{\pico\second} and the repetition rate of \SI{76}{\mega\hertz}, is split into two optical paths with horizontal- and vertical-polarization by \SI{40}{\milli\meter}-long calcite beam displacer. The pump light passes through a bandpass filter (\SI{940}{\nano\meter}, FWHM = \SI{10}{\nano\meter}) and enters the Sagnac-loop. A \SI{20}{\centi\meter}-long PMF is used as the optical nonlinear medium. The PMF is twisted by \SI{90}{\degree} such that its slow axes of each end are aligned to two counter-propagating pump polarizations. Correlated photon-pairs are generated at wavelengths of \SI{764}{\nano\meter} and \SI{1221}{\nano\meter} via SFWM, and the phase-matching condition is assisted by the birefringence of slow- and fast-axis of PMF \cite{CommercialFiberSagnac, PairGeneration_PMF_Walmsley}. The PMF does not have any active temperature stabilization. The generated photons are reflected by the bandpass filter centered at the pump's wavelength, and then recombined into a single spatial mode by the identical calcite beam displacer. The signal and idler photons are separated by a dichroic mirror, coupled into single-mode fibers, and then detected by a Silicon-based single-photon detector and superconducting nanowire single-photon detector, respectively. The polarization-correlations are analyzed by half-waveplates and linear polarizers. The pair production rate and heralding efficiencies are characterized by single and coincidence counting rates.

As shown in Table.~\ref{table:Walk-Off}, we estimated the spatial walk-off for signal and idler photon to be \SI{0.07}{\milli\meter} and \SI{0.09}{\milli\meter}, respectively. We model the photon's spatial modes in clockwise and counterclockwise paths as two Gaussian beams $F_{1}(x,y)$ and $F_{2}(x,y)$ with the beam diameter ($1/\text{e}$) of \SI{1.1}{\milli\meter}. The corresponding spatial amplitude overlap factor $\int F_{1}\cdot F_{2}\,dx\,dy$ between the two modes is calculated to be higher than \SI{99}{\percent} for both signal and idler wavelengths. Therefore, in this experiment, it is expected that the fiber-coupling efficiency drop due to the spatial walk-off is negligible. On the other hand, from the measured spectral full width at half maximum (FWHM) of the signal (\SI{0.8}{\nano\meter}) and idler photons (\SI{2.0}{\nano\meter}), we model the photon's amplitude temporal wavepacket with a Gaussian function whose temporal width ($1/\text{e}$) is \SI{1.3}{\pico\second} for both signal and idler photons. Similarly, the amplitude overlap factors between the wavepackets are calculated to be \SI{97.3}{\percent} and \SI{99.8}{\percent} for the signal and idler photons, respectively. Therefore, in this experiment, we introduced a \SI{1.6}{\centi\meter}--long quartz crystal in the signal arm to compensate for the temporal walk-off of signal photons.

\section{Result}

\begin{figure}[t]
\centering
\includegraphics[width=0.9\columnwidth]{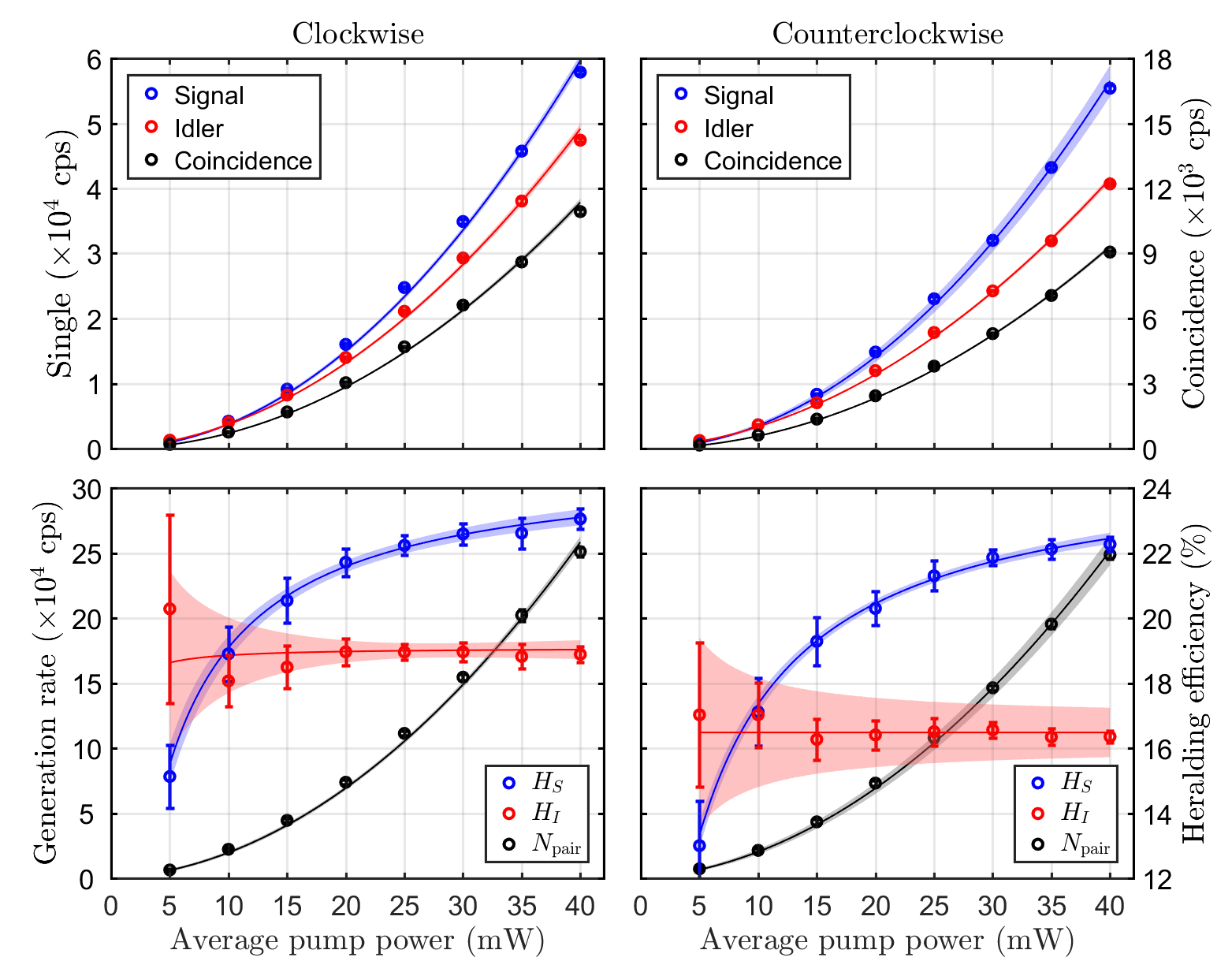}
\caption{Photon-pair generation rate for the clockwise (left) and counterclockwise (right) paths of the Sagnac-loop. The circles are the mean values of ten separate measurements of single and coincidence counting rates with the standard deviations as the error bars. The solid curves are fits to the model~\ref{eq:CountingModel}, and the shaded regions represent the \SI{95}{\percent} confidence interval for the fitting curve.}
\label{fig:CountRate}
\end{figure}

\subsection{Correlated photon-pair generation from a polarization-maintaining optical fiber}\label{sec:countingrate}

We measure the single and coincidence counting rates for the clockwise and counterclockwise paths with a coincidence window $\Delta t_{C}$ of \SI{1}{\nano\second}, as shown in figure~\ref{fig:CountRate}. After subtracting the detector's dark counts from the single counting rate, we calculate the mean values for the single and coincidence counting rates over five separate measurements. The measured single $N_{S(I)}$ and coincidence $N_{C}$ counting rates are used to estimate the heralding efficiencies, $H_{S(I)} = N_{C}/N_{I(S)}$, and the pair generation rate, $N_{\text{Pair}}=\left(N_{S}N_{I}\right)/N_{C}$. The quadratic scaling of the counting rates with the incident pump power shows that the pair production process is governed by the third-order optical nonlinear process. Therefore, it is rational to quantify the pair generation rate per pulse in units of cps$/$mW$^{2}$, which is estimated to be $2.1(2) \times 10^{-6}$ cps/mW$^{2}$ per pulse for our experiment.

In figure~\ref{fig:CountRate}, it is notable that the heralding efficiency for the signal photons varies as a function of pump power while that of the idler photons stays relatively constant. This feature is not typically observed in the spontaneous parametric down-conversion process. To study this behavior, we modeled the single and coincidence rate with the inclusion of the Raman-scattered photons from the PMF. In particular, we assumed that the incoherent Raman-scattering plays a major role as background noise in our experiment which scales linearly with pump power \cite{RamanNoise}. The modeled single and coincidence counting rates are expressed as

\begin{eqnarray}
N_{S} &= \eta_{S} N_{\text{Pair}} P^{2} + N_{D,S} + \eta_{S} N_{\text{Raman,S}} P, \\
N_{I} &= \eta_{I} N_{\text{Pair}} P^{2} + N_{D,I} + \eta_{I} N_{\text{Raman,I}} P, \\
N_{C} &= \eta_{S} \eta_{I} N_{\text{Pair}} P^{2} \\
&+ \eta_{S} N_{\text{Pair}} P^{2} (N_{D,I} + N_{\text{Raman,I}} P \eta_{I}) \Delta t_{C} \\
&+ \eta_{I} N_{\text{Pair}} P^{2} (N_{D,S} + N_{\text{Raman,S}} P \eta_{S}) \Delta t_{C} \\
&+ (N_{D,I} + \eta_{I} N_{\text{Raman,I}} P )(N_{D,S} + \eta_{S} N_{\text{Raman,S}} P ) \Delta t_{C},
\label{eq:CountingModel}
\end{eqnarray}
 where $\eta_{S(I)}$ is the total detection efficiencies of signal(idler) photons including fiber-coupling efficiency, photon loss by imperfect optics, and detection efficiency. $N_{\text{Pair}}$ is the pair generation rate of SFWM process, $N_{\text{Raman,S(I)}}$ is the Raman-scattering rates at the signal(idler) wavelengths, $N_{D,S(I)}$ is the dark counts for the signal(idler) photon detectors, and $P$ is the average pump power.

We applied the least-squares method to perform a global fitting of the model~\ref{eq:CountingModel} to the experimental data of \{$N_{S},N_{I},N_{C},H_{S},H_{I}$\} with the shared fitting parameters of \{$N_{\text{Pair}}$, $N_{\text{Raman,S}}$, $N_{\text{Raman,I}}$, $\eta_{S}$, $\eta_{I}$\}. Then, we used the optimized parameters to estimate the pair generation rate $N_{S}N_{I}/N_{C}$, as shown in figure~\ref{fig:CountRate}. In figure~\ref{fig:CountRate}, the symbols represent the experimentally obtained values from the measured single and coincidence count rate after the subtraction of the dark counts of each detector. The solid lines are a curve fit based on the model and the shaded regions represent the \SI{95}{\percent} confidence interval for the fitting. The fit parameters are obtained to be \{$N_{\text{Pair}}=149.71$,  $N_{\text{Raman,S}}=18.61$, $N_{\text{Raman,I}}=450.39$, $\eta_{S}=0.25$, $\eta_{I}=0.19$\}, which agrees with the fact that the incoherent Raman-scattering dominates in longer wavelength. From our model, we interpret that the varying heralding efficiency is due to the fact that the incoherent Raman noise dominates in idler arm for the low pump power regime. Thus, a click on the idler detector has a low probability of heralding the presence of a photon in the signal arm.

\begin{figure}[t]
\centering
\includegraphics[width=0.8\columnwidth]{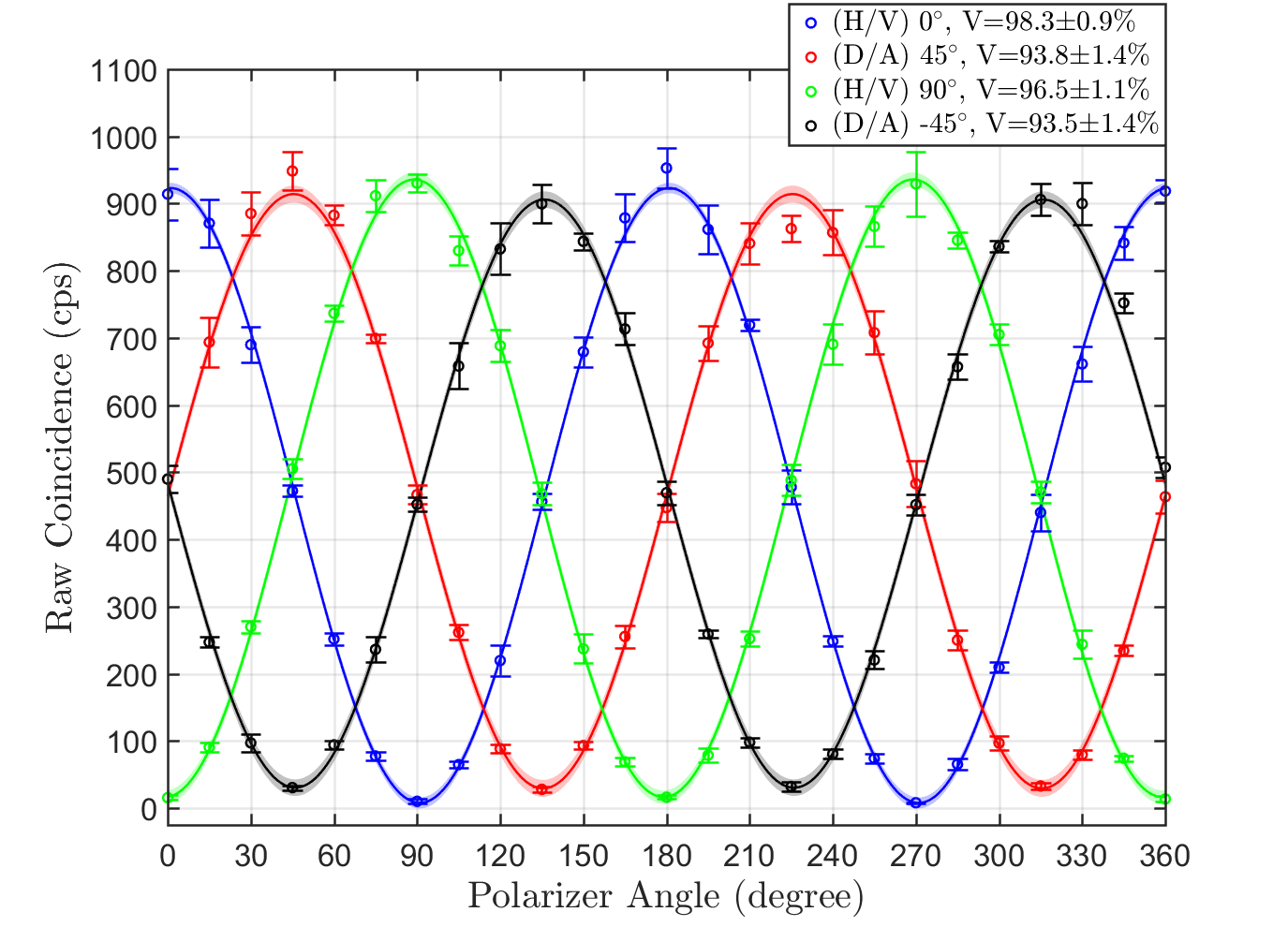}
\caption{Measurement of polarization entanglement: the circles are experimental data with the standard deviation calculated over five separate measurements. The solid curves are sinusoidal fits, and the shaded regions represent the \SI{95}{\percent} confidence interval for the sinusoidal fits. The visibilities are calculated from raw coincidence counts without any background subtraction.}
\label{fig:Pol_Entangle}
\end{figure}

\subsection{Polarization-entangled photon-pairs}

Polarization entanglement produced by the BD-Sagnac-EPS was characterized by performing polarization-correlation measurements. We measured the coincidence counting rates as a function of the rotation angle of a polarizer in the signal arm while the polarization measurement bases for the idler photons are fixed to horizontal ($H$, \SI{0}{\degree}), vertical ($V$, +\SI{90}{\degree}), diagonal ($D$, +\SI{45}{\degree}), and antidiagonal ($A$, -\SI{45}{\degree}), as shown in figure~\ref{fig:Pol_Entangle}. The measurement was performed under the average pump power of \SI{20}{\milli\watt} with a coincidence window of \SI{0.6}{\nano\second}. For each polarization setting, the mean values and standard deviations are calculated over five separate measurements of the single and coincidence counting rates. During these measurements, the single counting rates of the signal and idler photons remained constant at around \SI{13,000}{cps} and \SI{8,000}{cps}, respectively, during all four polarization-correlation measurements. A linear least-squares method was applied to perform a sinusoidal fit to the 2$\pi$-rotation of polarization-correlation measurement. 

The visibilities, $\text{V} = \left(\text{max}\{N_{C}\}-\text{min}\{N_{C}\}\right)/\left(\text{max}\{N_{C}\}+\text{min}\{N_{C}\}\right)$, calculated from raw coincidence counting rates for the idler's polarization bases at H, V, D, A are \SI{98.3(9)}{\percent}, \SI{96.5(11)}{\percent}, \SI{93.8(14)}{\percent}, and \SI{93.5(14)}{\percent}, respectively with the averaged value of $V_{\text{avg}} = \SI{95.5(6)}{\percent}$. We also tested the CHSH-inequality \cite{CHSHinequality} with the measured coincidence rates in figure~\ref{fig:Pol_Entangle}, and obtained the parameter $S=2.70\pm0.04$ which is in good agreement with the expected experimental value $S_{\text{exp}} = 2\sqrt{2}V_{\text{avg}}$. The strong violation of the CHSH-inequality by 17.5 standard deviations is a direct indication of the polarization entanglement. Based on the above model for counting rates together with the analysis on the spatial and temporal walk-off, the reduced entanglement quality was mainly attributed to spontaneous Raman scattering in PMF at idler wavelength and the uncompensated temporal walk-off for the idler photons.

\begin{figure}[t]
\centering
\includegraphics[width=1\columnwidth]{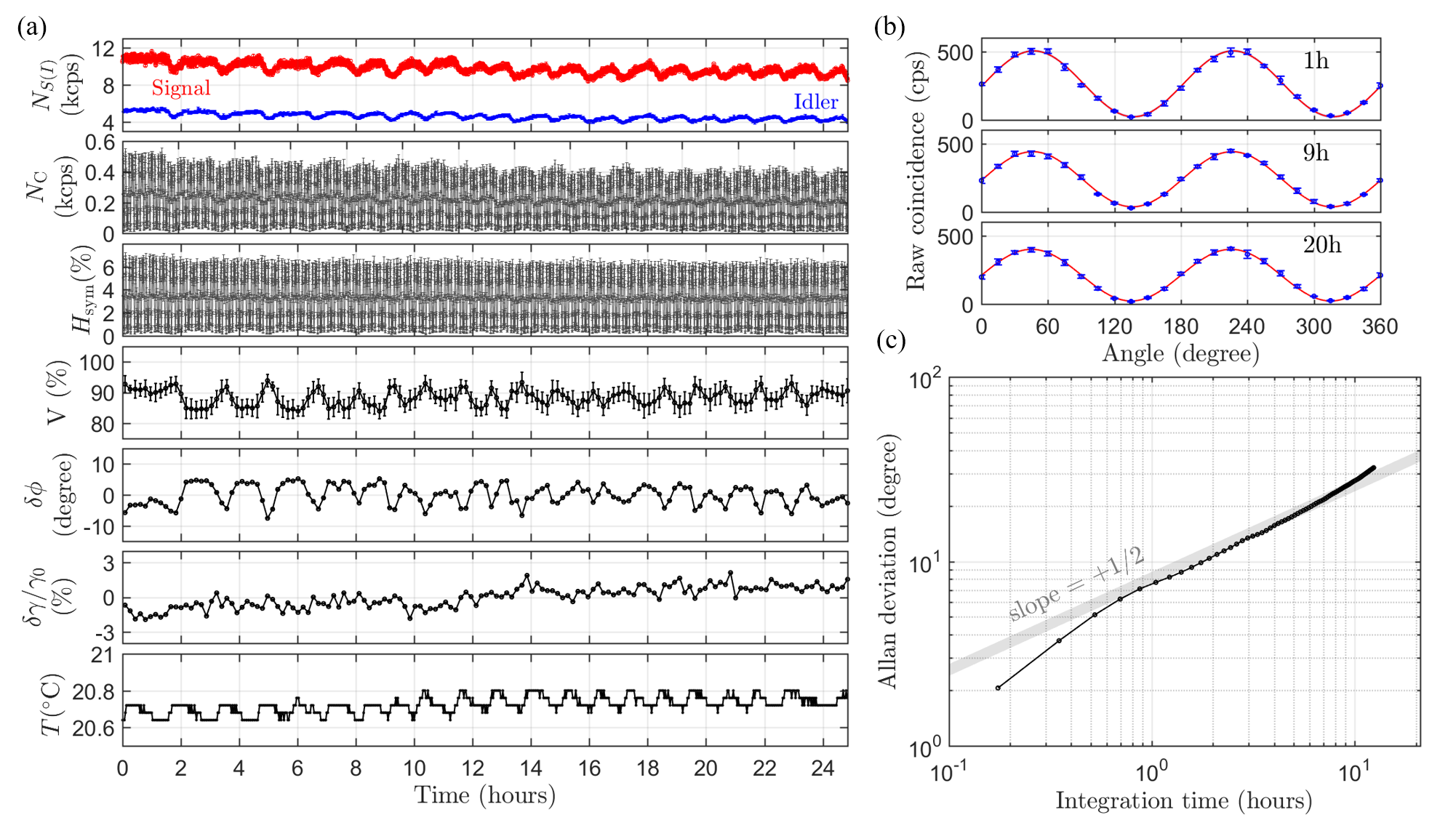}
\caption{The phase-stability of the BD-Sagnac-EPS. (a) Polarization-correlation measurement for over 24 hours. Single and coincidence counting rates are measured as a function of the rotation angle of the polarizer at signal arm while the polarization basis of the idler photons is fixed to the diagonal state (D, +\SI{45}{\degree}). The visibility V, the variation of the phase $\delta\phi$, and the relative amplitude fluctuation $\delta\gamma/\gamma_{0}$ ($\gamma_{0}=1/\sqrt{2}$) are extracted by fitting the model~\ref{eq:pol_corr_model} to the symmetric heralding efficiency. The temperature of our laboratory $T$ was monitored during the measurement (see text for a detailed analysis). (b) The magnified plot of raw coincidence counts for different measurement times. (c) Allan phase deviation as a function of the integration time. Block dots denote experimental data and a gray area represents lines with the slope +1/2.}
\label{fig:PhaseStability}
\end{figure}

\subsection{Phase-stability}

In order to characterize the phase-stability of the BD-Sagnac-EPS, we performed the polarization-correlation measurement for over 24 hours without active phase-stabilization, as shown in figure~\ref{fig:PhaseStability}. The temperature of the laboratory environment was regulated within a peak-to-peak value of \SI{0.16}{\degree}C. We performed a full 2$\pi$-rotation of the signal polarization basis at a period of ten minutes while the idler polarization basis is fixed to the diagonal basis ($D$, +\SI{45}{\degree}). The coincidence measurements are averaged over five separate measurement events at each rotation angle setting. We observed that the variation of the visibility V remains within the range of \SIrange{83.9}{93.9}{\percent}, as shown in figure~\ref{fig:PhaseStability}(a).

To extract the phase information from the measured data, we model the coincidence rate $N_{C}$ in terms of the phase and amplitude of the produced entangled state. We assume that the entangled state is a pure state $\ket{\Phi}=\gamma\ket{HH}+\sqrt{1-\gamma^2} e^{i\phi}\ket{VV}$ and ignore the contributions from the Raman-scattering and the detector's dark counts. With the idler polarization basis fixed at a diagonal state $\ket{M_{S}} = 1/\sqrt{2}\bigl(\ket{H}+\ket{V}\bigr)$ while rotating the signal basis $\ket{M_{I}} = \cos\theta\ket{H}+\sin\theta\ket{V}$, the coincidence counting rate can be expressed as
\begin{eqnarray}
N_{C} &= N_{\Phi}\eta_{S}\eta_{I}|\bra{M_{S}M_{I}}\Phi\rangle|^{2}\\
 &= \frac{N_{\Phi}\eta_{S}\eta_{I}}{2}\bigl(\gamma^2\cos^{2}\theta+(1-\gamma^2)\sin^{2}\theta + \gamma\sqrt{1-\gamma^2}\sin2\theta\cos\phi\bigr).
\label{eq:pol_corr_model}
\end{eqnarray}
Here $N_{\Phi}$ denotes a photon pair production rate of the entangled state which equals the double of the pair production rate $N_{\text{Pair}}$ of SFWM discussed in section~\ref{sec:countingrate}. Note that single counting rates for signal and idler are constant values $N_{S(I)}=\eta_{S(I)}N_{\Phi}/2$ over a full 2$\pi$-rotation of the signal polarization basis. The amplitude $\gamma$ and phase $\phi$ can be replaced with their mean values and their fluctuation; $\gamma \rightarrow 1/\sqrt{2} + \delta\gamma$ and $\phi\rightarrow \delta\phi$. Then, the second-order Taylor expansion simplifies the above expression~\ref{eq:pol_corr_model} as
\begin{equation}
N_{C}\approx N_{\Phi}\eta_{S}\eta_{I}\frac{1}{4}\bigl[1+\sin(2\theta+\sin^{-1}\bigl(2\sqrt{2}\delta\gamma)\bigr)\cos\delta\phi\bigr].
\label{eq:approx_pol_corr_model}
\end{equation}

The amplitude $\delta\gamma$ and phase fluctuation $\delta\phi$ appear in the polarization-correlation measurement as a phase-shift and visibility drop, respectively. Therefore, one can extract them by fitting the model to the measured coincidences with the fitting parameters of $\gamma$ and $\phi$. However, the coincidence counting rate $N_{C}$ depends not only the systematic errors $\gamma$ and $\phi$ of the entangled state caused by the instability of the interferometer, but it also reflects the instability of the other instrument such as the pump laser, i.e., the pair generation rate $N_{\Phi}$. In fact, we observed considerable fluctuations in single counting rates, as shown in figure~\ref{fig:PhaseStability}(a). Since our interest is the phase-stability of the designed interferometer, we take a symmetrized heralding efficiency $H_{\text{sym}} = N_{C}/\sqrt{N_{S}N_{I}}$ as a more appropriate parameter that is not affected by the fluctuating pair generation rate. Then, we fit the model to the heralding efficiency calculated from the raw single and coincidence rates for every full rotation of the signal polarizer, and obtain 143 values of the amplitude $\gamma$ and phase $\phi$ of the entangled state, as shown in figure~\ref{fig:PhaseStability}(a). 

In figure~\ref{fig:PhaseStability}(a), we observed that the standard deviation of the phase fluctuation is \SI{3.17}{\degree} under the temperature variation of \SI{0.04}{\degree}C, which is ten times larger than our estimation in section~\ref{sec:concept_design}. Since the robustness of the BD-Sagnac-EPS to the temperature variation is inherently obtained from its geometric Sagnac-configuration, we stress that this unexpected instability does not originate from the intrinsic property of the design as well as the components used. We believe that the instability may be mainly attributed to a combination of multiple external reasons, e.g., pump wavelength drift and a temperature-dependent phase-matching condition in PMF. For example, figure~\ref{fig:PhaseStability}(a) shows that the visibility variation is correlated with the single counting rates. One possible explanation for this is that the spectral instability of our mode-locked pump laser causes a slight wavelength-shift which may impact the relative phase $\phi(\lambda,T)$ of the entangled state. In our setup, the quartz crystal compensates only for the signal's temporal walk-off. In this case, the pump wavelength-dependent phase-shift is estimated to be \SI{1.01}{\radian}/\si{\nano\meter}~\footnote{If the temporal walk-off is compensated for both signal and idler photons, the pump wavelength-dependent phase-shift is estimated to be \SI{0.48}{\radian}/\si{\nano\meter}.}, which corresponds to the \SI{5.8}{\degree} phase-shift due to \SI{0.1}{\nano\meter} pump wavelength-shift. The origin of the instability will be further investigated in the future. 

To quantify the long-term phase-stability of the interferometer, we calculate the Allan deviation with the extracted phases $\phi$~\cite{AllanDev}. The Allan deviation with respect to the integration time $T$ is defined as 
\begin{equation}
\sigma(T=N\tau) = \sqrt{\frac{1}{2}\left<\left(\Delta\phi\right)^{2}\right>} = \sqrt{\frac{1}{2(N_{\text{tot}}-N)}\sum_{i=1}^{N_{\text{tot}}-N}(\phi_{i+N}-\phi_{i})^{2}}.
\label{AllanDeviation}
\end{equation}
The overall integration time, $T=N\tau$, is divided into $N$ samples with the time interval $\tau =$ 10 minutes and the total number of samples $N_\text{tot}=143$. We observed the Allan deviation of \SI{8}{\degree} over the integration time of 1 hour, as shown in figure~\ref{fig:PhaseStability}(c). Note that the optical path length from the first beam displacer to the second beam displacer in our setup is approximately \SI{1.3}{\meter}. The \SI{8}{\degree} phase uncertainty corresponds to the relative length deviation of $2.1 \times 10^{-8}$ per hour. This phase-stability is comparable with the results obtained from a typical Sagnac interferometer with strong laser light~\cite{MZAllan}. Finally, it is worth noting that the Allan deviation allows us not only to quantify the phase-stability but also to identify the source of noises by its scaling behavior \cite{PowerLaw}. In figure~\ref{fig:PhaseStability}(c), the measured Allan deviation follows the square-root scaling which indicates that the random-walk noise plays a major role in the long-term phase instability. 

\section{Conclusion}
We designed a novel interferometric configuration for polarization-entangled photon sources. By taking advantage of the intrinsic phase-stability and the utility in traditional Sagnac-type and Mach-Zehnder-type sources, our beam displacement Sagnac-interferometer is a practical and versatile source of entangled photons that can be implemented at various platforms without the requirement of customizing multi-wavelength polarization optics. We presented the detailed analysis for the spatiotemporal walk-off and the temperature-dependent phase-shift with commercially available beam displacers. The analysis showed that the designed interferometer is suitable for highly non-degenerate wavelengths and thermally stable. Furthermore, the alignment procedure for our scheme is more straightforward than traditional Sagnac-type sources.

We experimentally tested the functionality of the designed configuration with the pulsed generation of polarization-entangled photon-pairs at the wavelengths of \SI{764}{\nm} and \SI{1221}{\nm} from polarization-maintaining fiber via spontaneous four-wave mixing process. The setup consisted of only commercial off-the-shelf optical items. The strong violation of CHSH-inequality with parameter $S=2.70\pm0.04$ verified the polarization entanglement of the generated photon-pairs. The long-term phase-stability of the designed interferometer was characterized by performing the polarization correlation measurement over 24 hours, and during that time the visibility remains within a range of \SIrange{83.9}{93.9}{\percent} without active phase-stabilization. We attribute the observed instability mainly to the external instrumental imperfections, not to the intrinsic property of the designed interferometer or the used components, and this will be further investigated in the future. Nonetheless, the Allan deviation of less than \SI{8}{\degree} over the integration time of 1 hour quantitatively showed a good long-term phase-stability of the interferometer. Our prototype demonstration can be further optimized in the future, and we believe that this stable, practical, and versatile source is a promising candidate for both scientific research and industrial applications in quantum communication, sensing, and information processing. 

\section*{Funding}
This research was supported in part by the Canada Foundation for Innovation (25403, 30833); Ontario Research Foundation (098, RE08-051); Natural Sciences and Engineering Research Council of Canada (RGPIN-386329-2010); Industry Canada.

\ack{Y.S.L acknowledges support from the Mike and Ophelia Lazaridis Fellowship Program. M.X acknowledges support from Tsinghua University independent research program-International cooperation project (20173080024). R.T. acknowledges the personal support from the Alexander Graham Bell Canada Graduate Scholarship.}

\section*{References}

\bibliographystyle{unsrt}
\bibliography{draft_BD_sagnac_EPS}

\begin{thebibliography}{10}

\bibitem{QuantumEntanglement}
Ryszard Horodecki, Pawe\l{} Horodecki, Micha\l{} Horodecki, and Karol
  Horodecki.
\newblock Quantum entanglement.
\newblock {\em Rev. Mod. Phys.}, 81:865--942, Jun 2009.

\bibitem{Satellite_PKnight}
J~G Rarity, P~R Tapster, P~M Gorman, and P~Knight.
\newblock Ground to satellite secure key exchange using quantum cryptography.
\newblock {\em New Journal of Physics}, 4:82--82, oct 2002.

\bibitem{Satellite_AZeilinger}
M.~{Aspelmeyer}, T.~{Jennewein}, M.~{Pfennigbauer}, W.~R. {Leeb}, and
  A.~{Zeilinger}.
\newblock Long-distance quantum communication with entangled photons using
  satellites.
\newblock {\em IEEE Journal of Selected Topics in Quantum Electronics},
  9(6):1541--1551, 2003.

\bibitem{Satellite_ALing}
Robert Bedington, Juan~Miguel Arrazola, and Alexander Ling.
\newblock Progress in satellite quantum key distribution.
\newblock {\em npj Quantum Information}, 3(1):30, Aug 2017.

\bibitem{Compre_design_JP}
J.-P. Bourgoin, E.~Meyer-Scott, B.~L. Higgins, B.~Helou, C.~Erven,
  H.~H{\"u}bel, R.~Laflamme, T.~Jennewein, B.~Kumar, D.~Hudson, I.~D'Souza, and
  R.~Girard.
\newblock A comprehensive design and performance analysis of low earth orbit
  satellite quantum communication.
\newblock {\em New Journal of Physics}, 15(2):35, 2013.

\bibitem{FiberMichelson_PremKumar}
Chuang Liang, Kim~Fook Lee, Todd Levin, Jun Chen, and Prem Kumar.
\newblock Ultra stable all-fiber telecom-band entangled photon-pair source for
  turnkey quantum communication applications.
\newblock {\em Opt. Express}, 14(15):6936--6941, Jul 2006.

\bibitem{Sandwitch_AKarlsson}
Matthew Pelton, Philip Marsden, Daniel Ljunggren, Maria Tengner, Anders
  Karlsson, Anna Fragemann, Carlota Canalias, and Fredrik Laurell.
\newblock Bright, single-spatial-mode source of frequency non-degenerate,
  polarization-entangled photon pairs using periodically poled ktp.
\newblock {\em Opt. Express}, 12(15):3573--3580, Jul 2004.

\bibitem{Sandwitch_AZeilinger}
Hannes H\"{u}bel, Michael~R. Vanner, Thomas Lederer, Bibiane Blauensteiner,
  Thomas Lor\"{u}nser, Andreas Poppe, and Anton Zeilinger.
\newblock High-fidelity transmission of polarization encoded qubits from an
  entangled source over 100 km of fiber.
\newblock {\em Opt. Express}, 15(12):7853--7862, Jun 2007.

\bibitem{SagnacPeriscope_AZeilinger}
Michael Hentschel, Hannes H\"{u}bel, Andreas Poppe, and Anton Zeilinger.
\newblock Three-color sagnac source of polarization-entangled photon pairs.
\newblock {\em Opt. Express}, 17(25):23153--23159, Dec 2009.

\bibitem{UnfoldedSagnac_AKarlsson}
S.~Sauge, M.~Swillo, M.~Tengner, and A.~Karlsson.
\newblock A single-crystal source of path-polarization entangled photons at
  non-degenerate wavelengths.
\newblock {\em Opt. Express}, 16(13):9701--9707, Jun 2008.

\bibitem{Sagnac_Kim}
Taehyun Kim, Marco Fiorentino, and Franco N.~C. Wong.
\newblock Phase-stable source of polarization-entangled photons using a
  polarization sagnac interferometer.
\newblock {\em Phys. Rev. A}, 73:012316, Jan 2006.

\bibitem{DispersionShiftedFiberSagnac}
Xiaoying Li, Paul~L. Voss, Jay~E. Sharping, and Prem Kumar.
\newblock Optical-fiber source of polarization-entangled photons in the 1550 nm
  telecom band.
\newblock {\em Phys. Rev. Lett.}, 94:053601, Feb 2005.

\bibitem{CommercialFiberSagnac}
Bin Fang, Offir Cohen, and Virginia~O. Lorenz.
\newblock Polarization-entangled photon-pair generation in commercial-grade
  polarization-maintaining fiber.
\newblock {\em J. Opt. Soc. Am. B}, 31(2):277--281, Feb 2014.

\bibitem{MicrostructureFiberSagnac}
J.~Fan, M.~D. Eisaman, and A.~Migdall.
\newblock Bright phase-stable broadband fiber-based source of
  polarization-entangled photon pairs.
\newblock {\em Phys. Rev. A}, 76:043836, Oct 2007.

\bibitem{PhotonicCrystalFiberSagnac}
J\'er\'emie Fulconis, Olivier Alibart, Jeremy~L. O'Brien, William~J. Wadsworth,
  and John~G. Rarity.
\newblock Nonclassical interference and entanglement generation using a
  photonic crystal fiber pair photon source.
\newblock {\em Phys. Rev. Lett.}, 99:120501, Sep 2007.

\bibitem{PPLN-WG_Sagnac}
Han~Chuen Lim, Akio Yoshizawa, Hidemi Tsuchida, and Kazuro Kikuchi.
\newblock Stable source of high quality telecom-band polarization-entangled
  photon-pairs based on a single, pulse-pumped, short ppln waveguide.
\newblock {\em Opt. Express}, 16(17):12460--12468, Aug 2008.

\bibitem{PPLN-ridgeWG_Sagnac}
Shin Arahira, Naoto Namekata, Tadashi Kishimoto, Hiroki Yaegashi, and Shuichiro
  Inoue.
\newblock Generation of polarization entangled photon pairs at
  telecommunication wavelength using cascaded $\chi$(2) processes in a
  periodically poled linbo3 ridge waveguide.
\newblock {\em Opt. Express}, 19(17):16032--16043, Aug 2011.

\bibitem{PPLN-WG_Sagnac_fullfiber}
P.~{Vergyris}, F.~{Kaiser}, E.~{Gouzien}, G.~{Sauder}, T.~{Lunghi}, and
  S.~{Tanzilli}.
\newblock {Fully guided-wave photon pair source for quantum applications}.
\newblock {\em Quantum Science and Technology}, 2(2):024007, June 2017.

\bibitem{PPKTP-WG_Sagnac}
Evan Meyer-Scott, Nidhin Prasannan, Christof Eigner, Viktor Quiring, John~M.
  Donohue, Sonja Barkhofen, and Christine Silberhorn.
\newblock High-performance source of spectrally pure, polarization entangled
  photon pairs based on hybrid integrated-bulk optics.
\newblock {\em Opt. Express}, 26(25):32475--32490, Dec 2018.

\bibitem{Constraint_released_Sagnac}
Marco Fiorentino, Ga\'etan Messin, Christopher~E. Kuklewicz, Franco N.~C. Wong,
  and Jeffrey~H. Shapiro.
\newblock Generation of ultrabright tunable polarization entanglement without
  spatial, spectral, or temporal constraints.
\newblock {\em Phys. Rev. A}, 69:041801, Apr 2004.

\bibitem{Kurtisefer_RandomnessExtraction}
Lijiong Shen, Jianwei Lee, Le~Phuc Thinh, Jean-Daniel Bancal, Alessandro
  Cer\`e, Antia Lamas-Linares, Adriana Lita, Thomas Gerrits, Sae~Woo Nam,
  Valerio Scarani, and Christian Kurtsiefer.
\newblock Randomness extraction from bell violation with continuous parametric
  down-conversion.
\newblock {\em Phys. Rev. Lett.}, 121:150402, Oct 2018.

\bibitem{BD_Beausoleil}
Marco Fiorentino and Raymond~G. Beausoleil.
\newblock Compact sources of polarization-entangled photons.
\newblock {\em Opt. Express}, 16(24):20149--20156, Nov 2008.

\bibitem{BD_THumble}
P.~G. Evans, R.~S. Bennink, W.~P. Grice, T.~S. Humble, and J.~Schaake.
\newblock Bright source of spectrally uncorrelated polarization-entangled
  photons with nearly single-mode emission.
\newblock {\em Phys. Rev. Lett.}, 105:253601, Dec 2010.

\bibitem{LoopholeFreeTest}
Lynden~K. Shalm, Evan Meyer-Scott, Bradley~G. Christensen, Peter Bierhorst,
  Michael~A. Wayne, Martin~J. Stevens, Thomas Gerrits, Scott Glancy, Deny~R.
  Hamel, Michael~S. Allman, Kevin~J. Coakley, Shellee~D. Dyer, Carson Hodge,
  Adriana~E. Lita, Varun~B. Verma, Camilla Lambrocco, Edward Tortorici, Alan~L.
  Migdall, Yanbao Zhang, Daniel~R. Kumor, William~H. Farr, Francesco Marsili,
  Matthew~D. Shaw, Jeffrey~A. Stern, Carlos Abell\'an, Waldimar Amaya, Valerio
  Pruneri, Thomas Jennewein, Morgan~W. Mitchell, Paul~G. Kwiat, Joshua~C.
  Bienfang, Richard~P. Mirin, Emanuel Knill, and Sae~Woo Nam.
\newblock Strong loophole-free test of local realism.
\newblock {\em Phys. Rev. Lett.}, 115:250402, Dec 2015.

\bibitem{BD_RolfHorn}
Rolf Horn and Thomas Jennewein.
\newblock Auto-balancing and robust interferometer designs for polarization
  entangled photon sources.
\newblock {\em Opt. Express}, 27(12):17369--17376, Jun 2019.

\bibitem{SingaporeSource_ALing}
Alexander Lohrmann, Chithrabhanu Perumangatt, Aitor Villar, and Alexander Ling.
\newblock Broadband pumped polarization entangled photon-pair source in a
  linear beam displacement interferometer.
\newblock {\em Applied Physics Letters}, 116(2):021101, Jan 2020.

\bibitem{LoudonBook}
Rodney. Loudon.
\newblock {\em The quantum theory of light / Rodney Loudon}.
\newblock Clarendon Press Oxford, 1973.

\bibitem{BDformula}
M.~Avenda\ {n}o Alejo.
\newblock Analysis of the refraction of the extraordinary ray in a
  plane-parallel uniaxial plate with an arbitrary orientation of the optical
  axis.
\newblock {\em Opt. Express}, 13(7):2549--2555, Apr 2005.

\bibitem{book_PolarizationOptics}
J.N. Damask.
\newblock {\em Polarization Optics in Telecommunications}.
\newblock Springer Series in Optical Sciences. Springer.

\bibitem{book_Ghosh}
E.D. Palik and G.~Ghosh.
\newblock {\em Handbook of Optical Constants of Solids: Handbook of
  Thermo-Optic Coefficients of Optical Materials with Applications}.
\newblock Elsevier Science \& Technology Books, 1997.

\bibitem{thermo_optic_YVO4}
Yoichi Sato and Takunori Taira.
\newblock Highly accurate interferometric evaluation of thermal expansion and
  dn/dt of optical materials.
\newblock {\em Opt. Mater. Express}, 4(5):876--888, May 2014.

\bibitem{two-crystal-Sagnac}
Terence~E. Stuart, Joshua~A. Slater, F\'elix Bussi\`eres, and Wolfgang Tittel.
\newblock Flexible source of nondegenerate entangled photons based on a
  two-crystal sagnac interferometer.
\newblock {\em Phys. Rev. A}, 88:012301, Jul 2013.

\bibitem{PairGeneration_PMF_Walmsley}
Brian~J. Smith, P.~Mahou, Offir Cohen, J.~S. Lundeen, and I.~A. Walmsley.
\newblock Photon pair generation in birefringent optical fibers.
\newblock {\em Opt. Express}, 17(26):23589--23602, Dec 2009.

\bibitem{RamanNoise}
Robert~W. Boyd.
\newblock {\em Nonlinear Optics, Third Edition}.
\newblock Academic Press, Inc., USA, 3rd edition, 2008.

\bibitem{CHSHinequality}
John~F. Clauser, Michael~A. Horne, Abner Shimony, and Richard~A. Holt.
\newblock Proposed experiment to test local hidden-variable theories.
\newblock {\em Phys. Rev. Lett.}, 23:880--884, Oct 1969.

\bibitem{AllanDev}
D.~W. {Allan}.
\newblock Statistics of atomic frequency standards.
\newblock {\em Proceedings of the IEEE}, 54(2):221--230, 1966.

\bibitem{MZAllan}
Michal Mičuda, Ester Doláková, Ivo Straka, Martina Miková, Miloslav Dušek,
  Jaromír Fiurášek, and Miroslav Ježek.
\newblock Highly stable polarization independent mach-zehnder interferometer.
\newblock {\em Review of Scientific Instruments}, 85(8):083103, 2014.

\bibitem{PowerLaw}
WJ~Riley.
\newblock Handbook of frequency stability analysis.
\newblock {\em NIST}, 1065:1--123, 01 2007.

\end{thebibliography}
\end{document}